# Suppressing Kirkendall Void Density in Circuit Interconnections by Strain Annealing

Chongyang Cai,[a,b] Rong An,[a,b*] Chunqing Wang,[a,b] Yanhong Tian,[a,b]

[a] State Key Laboratory of Advanced Welding and Joining, Harbin Institute of Technology, Harbin 150001, China

[b] Micro- and Nanotechnology Research Center, Harbin Institute of Technology, Harbin 150080, China

*Corresponding author. Tel./fax:+86 451 86418725/86416186

E-mail address: anr@hit.edu.cn (Rong An).

**Abstract**

Unpredictable Kirkendall void formation at the interface of circuit interconnections underlies degradation in electronics, yet there is a lack of effective approaches to curb the amount of these voids. Here we developed a strain-anneal method to tailor grain size distributions in the copper substrate of interconnections, and demonstrate quantitatively that not only the removal of the impurities but also an increase in the grain size of the substrates leads to an appreciable decline in the void density. The interconnections on the substrate recrystallized at a high annealing temperature show the massive porosity and the increased sensitivity of the voiding to the grain size. Our findings have broad implications for manipulation of void propensity in many other hetero-interfaces and are essential for high-performance circuit bonding in high temperature/high power electronic devices based on wide band gap semiconductors.

## 1. Introduction

The prevalence of voids forming at the interface of circuit interconnections is becoming an increasing concern, especially for electronics under rapid scaling or harsh conditions or with increased service life [1]. These voids may pose a challenge to the long term stability of electronic devices due to their weakening effects on the mechanic and electronic properties of circuit interconnections. Moreover, they seem largely unpredictable and may cause crack propagation and thus reduce interconnection life severely by mere accident [2]. These voids, appearing within the interfacial intermetallic compound (IMC) layers at interconnections, are commonly referred as Kirkendall voids, which can be thought of as a consequence of the difference in diffusion rates of the metal atoms [3,4]. Nevertheless, a same sort of interconnections often exhibit inherently contradictory tendencies for the void amounts, even though the interconnections were identical in structure and composition, which indicates that this void formation is deeply influenced by other factors.

For sound circuit interconnections, there are two major built-in factors which determine the voiding propensity: solder and substrate. For example, certain alloying elements like Bi in solder can promote the formation of the voids [5]. This problem can be tackled effectively by changing the chemical composition of the used solder to avoid such elements. Another ongoing challenge arises from the substrate. Generally, electroplated copper, a commonly used substrate, tends to indicate high amounts of the interfacial voids. Indeed, the void nucleation is strongly facilitated by remained contaminants like sulfur or organic impurities derived from the electroplating bath [6]. To a certain extent, solder interconnects with a small number of interfacial voids can be produced by optimizing electroplating parameters [7]. Yet, unexpectedly, some Cu samples showed no correlation of the voiding tendency with electroplating performance and, additionally, some of the worst voiding occurred in the interconnections on the Cu pads prepared by high-end techniques for electroplating [8]. This suggests that other features in Cu substrates such as grain size may play a rather dominant role in such cases.

Cu pads, different in grain size, have been found to exhibit noticeable differences in tendency for void formation at the corresponding solder interconnects [9]. All these results have been obtained from interconnections on various Cu substrates (electrodeposited Cu, high purity Cu, vacuum sputtered Cu, etc.) fabricated by a process with different parameters, or even by different methods [10]; a complex interplay, however, exists between the grain size and impurity amount of these pads. Hence, it is hard to draw firm quantitative conclusions about the correlation between the substrate property and the void quantity.

In this paper, we introduce a modified strain-anneal technique combining with surface indentation for inhomogeneous recrystallization in copper substrates, to confirm a distinct role played by the substrate grain size in the void formation of circuit interconnections. Via critical strain-annealing, about 20-μm average grain size was obtained in an electrodeposited copper foil, and thus a substantial decline in the void density of circuit interconnections was achieved.

## 2. Materials and Methods

Two kinds of commercially available copper, high purity oxygen free copper (HPOFC, 99.9999 wt%) and electrolytic tough pitch copper (ETPC, 99.9 wt%), which were supplied by Unirise Special Copper Corporation, were used as substrates to prepare sample joints. These copper substrates (10 mm × 10 mm × 1 mm) were indented by a diamond cone indenter using a Rockwell hardness tester according to ASTM E18 [11] and then annealed at 600 °C or 700 °C (100-200 °C above its recrystallization temperature) for different time. The substrates were slightly etched by an etchant (5 mL, 30% $H_2O_2$ + 50 mL, 25% $NH_3 \cdot H_2O$) to reveal grain boundaries in copper.

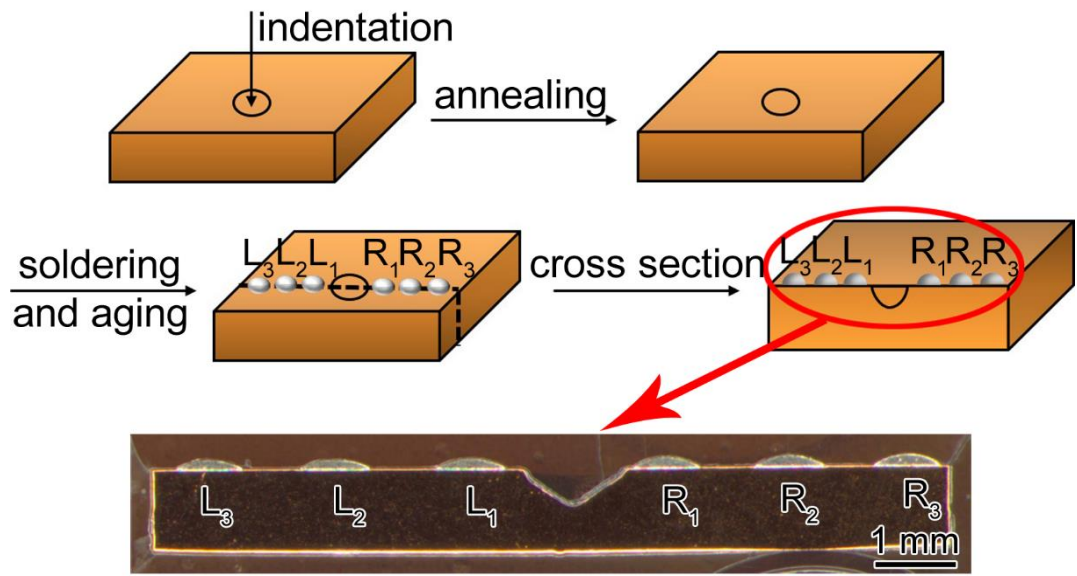

**Fig. 1.** Schematic illustration of the experimental design. The solder joints on the substrates were categorized into two groups, left side (L) ones and right side (R) ones; the subscript numbers are arranged in the order of increasing distance from the indentation.

Solder balls of eutectic Sn37Pb (400 μm diameter), which were provided by Indium Corporation, were subsequently reflowed on annealed substrates at 260 °C for 2 min.

The obtained solder joints were isothermally aged at 150 °C for different time. All the joints on the substrates then were sectioned and grounded; they were freshly polished before each quasi in-situ microstructural observation [12]. The schematic experimental configuration and resulting cross-sectional structures are shown in Fig. 1. An optical microscopy was employed to observe the grain boundary of copper substrates. A scanning electron microscope (SEM) using a backscattered electron signal was used for the imaging and analysis of the cross-sectional microstructures of the joints. Image analysis was carried out using ImageJ software.

## 3. Results and discussions

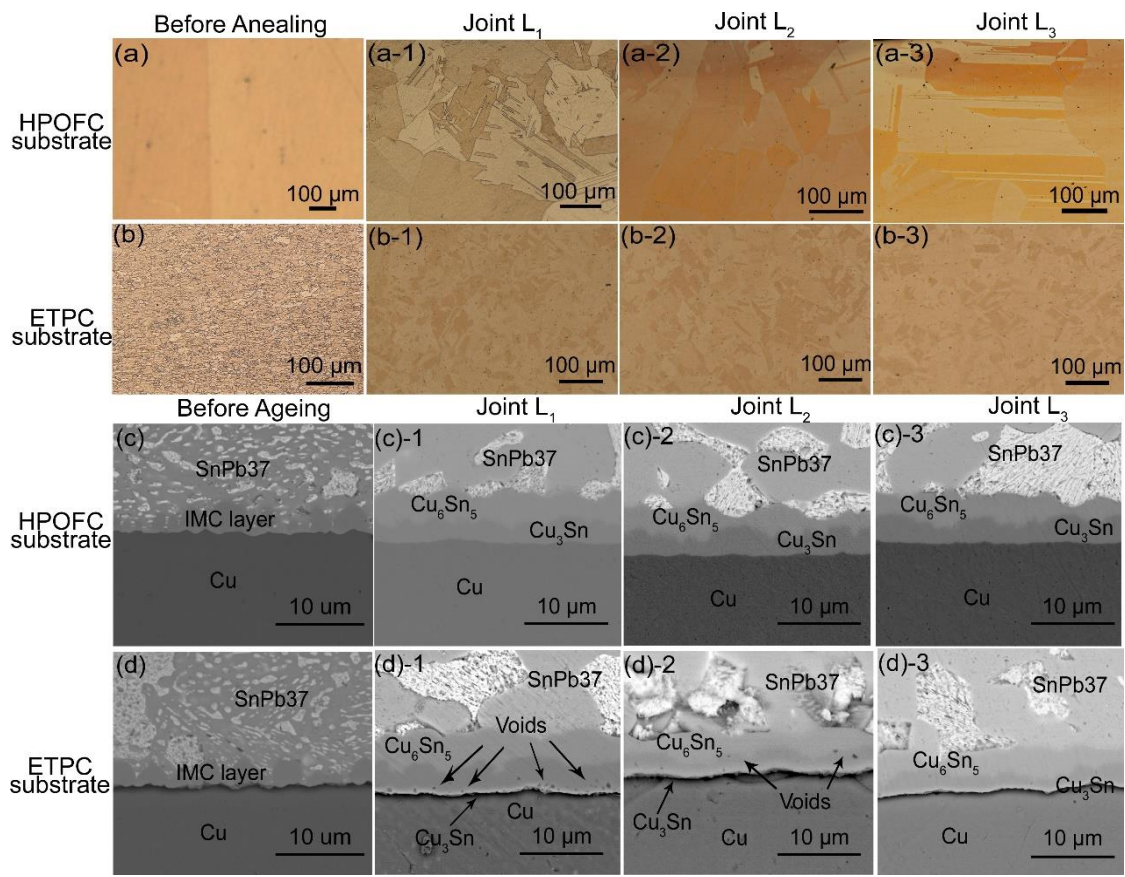

**Fig. 2.** Images showing cross-sectional structures of (a) HPOFC and (b) EPTC under the different joints before and after being annealed at 600 °C for 30 min. SEM micrographs of joints on (c) HPOFC and (d) EPTC substrate before and after being aged at 150 °C for 100 h.

Fig. 2 (a) and (b) show the optical metallographs of the etched copper substrates of left side joints. After annealing, large columnar crystal grains (average grain size ~2 mm) disappear and fine grains (average grain size ~50 μm) nucleate in the deformed regions near the indentation on HPOFC substrate. However, ETPC substrate bucks the trend: the annealed grains (average grain size ~20 μm) become even larger than the original (average grain size ~300 nm). Note also that in the as-received state the grain size of ETPC substrate is much smaller than HPOFC substrate, whereas after annealing both the substrates become nearly the same in grain size.

It is observed that continuous recrystallization occurs during the annealing. The new crystals can nucleate and grow in a certain plastically deformed region with residual stress. Once the deformed matrix is consumed by these new crystals, further annealing only causes grain growth [13]. The HPOFC substrate with minor impurities and crystal imperfections possesses a low density of potential nucleation sites for recrystallization during annealing. In this case, the fine recrystallized grains nucleate by consuming the virtually full amount of stored energy of the cold work, thus growing up slightly, and eventually making the resultant grains much smaller than the original [14].

For ETPC substrate, however, things are a bit different. Its recrystallization rate is high and the annealing time is enough for the grains to be fully recrystallized [15]. As a result, the time for the grain growth account for a bigger slice of the whole recrystallization period than the nucleation. That is the reason why its grain sizes after annealing are comparable to or much larger than the original ones.

When it comes to the variation in microstructures of the copper, both kinds of the Cu substrates show that the grain size increases significantly with decreasing strain in the direction away from the indentation. A grain size gradient is established around the surface indentation on the substrate as a result of recrystallization after annealing. Such trend is more discernible for HPOFC substrate. The grains under joint $L_1$ are generally the biggest in size; those under joint $L_2$ the second in size; and those under joint $L_3$ the smallest. Joint $L_1$ lies adjacent to the indentation and thus the grains beneath it store the largest amount of deformation energy. Therefore, they have more driving force for recrystallization than those grains under other joints [16].

The interface microstructures of different joints are shown in Fig. 2 (c) and (d). After being aged at 150 °C for 100 h, the scallop-like $Cu_6Sn_5$ flattens and IMC layer becomes thicker. All these features indicate that at 150 °C, the thermal aging duration is long enough for the elements to diffuse sufficiently to reveal appreciably detectable microstructural change.

After aging, almost no voids were detected in any joints on HPOFC substrate, whereas the interface at ETPC substrate has a considerable amount of voids, as shown in Fig. 2 (c) and (d). This difference can be mainly attributed to two reasons. First, ETPC substrate may still contain some residual impurities at the interface or grain boundaries even after completion of high temperature annealing treatments due to high levels of incorporated impurities derived from electroplating. These electroplating impurities can reduce the activation energy barrier for void nucleation and thus promote the formation of voids [17]. Second, as mentioned above, the EPTC substrate has much smaller grains than the HPOFC substrate and therefore has more grain boundaries for copper atoms to diffuse rapidly [18].

Fig. 2 (d) shows that the interfacial void amounts for EPTC substrate decrease with the increasing subscript number of the joints, which suggests a significant relation of the void propensity to the grain size of Cu. To perform more quantitative analysis, herein we introduce two physical quantities, $N$ grain size number and $D$ void density. $N$ denotes average grain size, and can be obtained using an intercept method recommended by the ASTM standard E112−12 [19]. It was calculated by

$$N = 10 - 2\log_2 \bar{l} \tag{1}$$

Where $\bar{l}$ is the average lineal intercept length. According to Equation (1), small grains usually correspond to large $N$. The other notion, $D$, was used to evaluate voiding propensity. $D$ is determined by summing the areas of all the voids and dividing it by the total length of IMC layer. Accordingly, $D$ has a positive correlation with the voiding propensity.

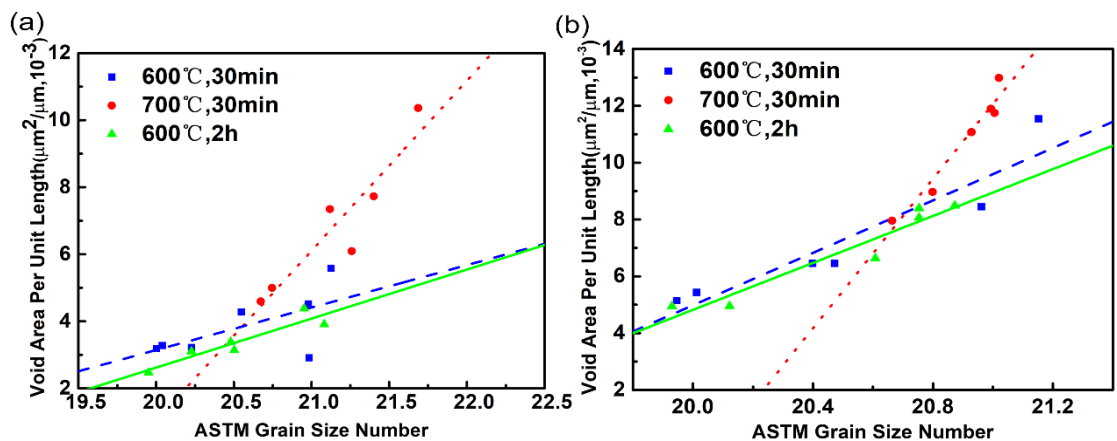

**Fig. 3.** Grain size number and void density of solder joints on ETPC substrate after being aged at 150 °C for (a) 100 h and (b) 200 h. ETPC substrates were annealed at 600 °C for 30 min and 2 h and annealed at 700 °C for 30 min.

The relationship between void density and grain size number after aging for 100 h is presented in Fig. 3 (a). Three sets of annealing parameters were examined for ETPC substrate to provide extensive data. In all these situations, both the joints $L_1$ and $R_1$, which are the closest to the indentation, are the largest in $N$ and $D$, and their coordinates are located on the top right side in Fig. 3 (a) or (b) accordingly. Moreover, the values of $N$ and $D$ decrease with increasing distance from the indentation to the joint. Indeed, irrespective of the annealing and aging parameters, $D$ is an essentially monotonic function of $N$ (Fig. 3). Using the method of least squares, we fit the data to a clear linear law shown by the dashed lines: $D \sim sN$, where s is the slope. Since $N$ is inversely proportional to grain size, the observed linear law behavior suggests that smaller grain sizes can be beneficial to the formation and coalescence of voids at the $Cu_3Sn$/ETPC interface.

When the aging time was prolonged to 200 h, the copper under the joints does not exhibits a noticeable change in grain size, maintaining an approximately constant range of $N$, from 19.8 to 21.2. This suggests that the aging temperature of 150 °C is not high enough for the grains to recrystallize nor grow. In contrast, with the extension of aging time, the void amount increases significantly, which is illustrated by a substantial rise in $D$, from [2.0, 4.5] to [4.5, 9.0] 10-3μm (Fig. 3). This implies that the voiding propensity increases with the aging time. Such kind of situation is commonly clarified as a result of the continuing of diffusion. Additionally, the emergence of more voids is usually accompanied by the thickening of IMC layer [20], which provides more space for voids to grow. In spite of that, the trend of $D$ is still consistent with the trend of $N$.

So we can reach the conclusion that smaller grains can indeed promote the voiding propensity.

Considering we have managed fitting data into a linear law, it turns out that no matter under which kind of heat treatment, $N$ and $D$ all follow a linear distribution model. Besides, all these slopes are positive, which adds weight to the inference that smaller grains (with larger $N$ values) can boost the development of voids at the interface. The only difference is the value of each fitting line's slope: the slope of the substrate annealed at 700 °C is remarkably higher than the substrate annealed at 600 °C. However, the fitting lines of substrates annealed at 600 °C are almost parallel to each other regardless of the annealing time, making slopes approximately the same.

Many theories were proposed to explain such relationship. Previously, the corresponding change of void tendency according to grain size change was viewed as a result of combined effect of the following two mechanisms. First, the inclination of void formation is mainly influenced by the number of effective vacancies [21]. Smaller grains are beneficial to the generation of voids, as there are more effective vacancies diffusing and coalescing at the interface to form voids. Also, the finer the grains are, the more grain boundaries the substrate will have. Owing to the increase of diffusion paths the substrate provides, more vacancy fluxes are generated to facilitate voiding process [18].

Theories like "effective vacancies" [9] or "diffusion paths" [18] may to a certain point demonstrate the linkage that fine-grained copper substrate contributes to voiding at joints. The intrinsic reason underlying the quantitative relationship between $N$ and $D$ but nonetheless remains ambiguous. To understand how grain size influences the voiding propensity thoroughly, the whole process of voiding is schematically described in Fig. 4.

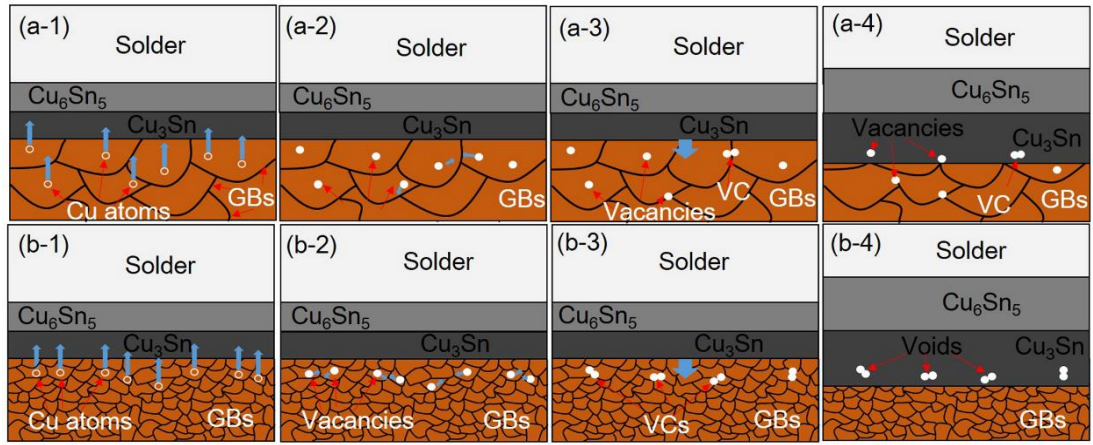

Fig. 4. Schematic diagrams describing the procedure of void formation. (a) Cu with large grains, and (b) Cu with small grains

During the aging period, copious amount of Cu atoms diffuse from Cu substrate to the interconnect interface due to the different diffusion coefficients between different metal atoms. The Cu flux diffuses to the other side, leaving remnant vacancies staying in the substrate. These residual vacancies can move towards grain boundaries and agglomerate into Vacancy Clusters (VCs) considering grain boundaries have an noticeable absorptive effect on vacancies, impurities and other defects [22]. Meanwhile, the IMC layer keeps on growing as diffusion proceeds. The result of IMC interface evolution is the interface of $Cu_3Sn$/Cu expanding towards the substrate [18]. With the invasion of $Cu_3Sn$, originally formed VCs are left behind, trapping inside $Cu_3Sn$ layer [22]. Finally, these VCs existing inside the IMC layer will further diffuse, aggregate and grow into visible voids under SEM images.

The experimental results are in good agreement with the theoretical model. Many factors concerning voiding propensity can be explained by this illustration. First, as shown in Fig. 2(d), most voids are found at the interface of $Cu_3Sn/Cu$ instead of $Cu_6Sn_5/Cu_3Sn$ because as described in the mechanism it is the border of $Cu_3Sn/Cu$ that expands towards substrate. Second, as shown in Fig. 4(b), once the coarse grains are replace by smaller size of grains in the substrate, the number of grain boundaries will increase markedly. With more grain boundaries to adsorb vacancies, there is a better chance for these voids to coalesce, as a result, more VCs are produced. Eventually it may lead to the result that more voids develop at the interface. Therefore the efficient way to keep the number of voids down is to ensure that the grains in substrates are large. What's more, the promotion of voids by impurities can also be explained. Since grain boundaries can adsorb vacancies as well as impurities, these impurities gathering around grain boundaries can act as nucleation centers and thus reduce the activation energy barrier for VCs to nucleate. The difference of fitting lines' slopes can also be interpreted. The slopes of joints being aged for 200 h are steeper than joints being aged for 100 h as more time is available for vacancy clusters to diffuse and emerge. Furthermore, it can be inferred that tougher heat treatment may promote further recrystallization and make grains finer after annealing while simply increasing annealing time will not do much help for the changing of grain distribution, thus can help explain the reason why fitting lines annealed at 600 °C resemble while fitting lines annealed at 700 °C are much steeper.

## 4. Conclusions

In summary, we have developed a modified strain-anneal method to achieve a grain size gradient spectrum in both HPOFC and ETPC substrate. This method is utterly effective for both kinds of substrates in achieving an architected spatial distribution of grain size. Microstructural observations show that the HPOFC grains become finer while ETPC grains are larger after being treated by surface indentation followed by strain-annealing yet no voids are found on the interface of HPOFC substrate. Effect of substrate grain size on the void propensity of the joints has been quantitatively investigated and indicates that the void density obey a linear distribution with the grain size number, and higher annealing temperature and longer aging time can bring more voids to come up, which can be attributed to both factors like grain size and the amount of residual impurities. Our results highlight an example of how grain size of substrates can be tailored to enable manipulation of void propensity in hetero-interfaces, and suggest a promising strategy for high-stability circuit bonding in electronic devices.

**Acknowledgements**

This work was supported by the National Natural Science Foundation of China [grant number 51005055] and the Fundamental Research Funds for the Central Universities [grant number HIT. NSRIF. 2015066].